# A superconducting nanowire photon number resolving four-quadrant detector-based Gigabit deep-space laser communication receiver prototype

## Author Information

Hao Hao[1], Qing-Yuan Zhao[1,2]✉, Yang-Hui Huang[1], Jie Deng[1], Hui Wang[1], Jia-Wei Guo[1], Shi Chen[1], Sai-Ying Ru[1], Zhen Liu[1], Yi-Jin Zhou[1], Shun-Hua Wang[1], Chao Wan[2], Hao Liu[2], Zhi-Jian Li[2], Hua-bing Wang[1,2], Xue-Cou Tu[1,3], La-Bao Zhang[1,3], Xiao-Qing Jia[1,3], Jian Chen[1,2], Lin Kang[1,3], and Pei-Heng Wu[1,3]

[1]*Research Institute of Superconductor Electronics (RISE), School of Electronic Science and Engineering, Nanjing University, Nanjing, Jiangsu 210023, China*

[2]*Purple Mountain Laboratories, Nanjing, Jiangsu 211111, China*

[3]*Hefei National Laboratory, Hefei, Anhui 230088, China*

✉ qyzhao@nju.edu.cn.

# Abstract

Deep space explorations require transferring huge amounts of data quickly from very distant targets. Laser communication is a promising technology that can offer a data rate of magnitude faster than conventional microwave communication due to the fundamentally narrow divergence of light. This study demonstrated a photon-sensitive receiver prototype with over Gigabit data rate, immunity to strong background photon noise, and simultaneous tracking ability. The advantages are inherited from a joint-optimized superconducting nanowire single-photon detector (SNSPD) array, designed into a four-quadrant structure with each quadrant capable of resolving six photons. Installed in a free-space coupled and low-vibration cryostat, the system detection efficiency reached 72.7%, the detector efficiency was 97.5%, and the total photon counting rate was 1.6 Gcps. Additionally, communication performance was tested for pulse position modulation (PPM) format. A series of signal processing methods were introduced to maximize the performance of the forward error correction (FEC) code. Consequently, the receiver exhibits a faster data rate and better sensitivity by about twofold (1.76 photons/bit at 800 Mbps and 3.40 photons/bit at 1.2 Gbps) compared to previously reported results (3.18 photon/bit at 622 Mbps for the Lunar Laser Communication Demonstration). Furthermore, communications in strong background noise and with simultaneous tracking ability were demonstrated aimed at the challenges of daylight operation and accurate tracking of dim beacon light in deep space scenarios.

# Introduction

Recent and upcoming deployments of satellite laser communication systems have superior high-speed data transmission[1]. Consequently, this has been demonstrated to be a revolution in communication, on Earth, between constellation satellites, and across the solar system[2]. Since laser wavelength is about $10^5$ times shorter than the microwave, laser communications benefit from the small divergence for lasers. The magnitude of signal power concentrated at the receiver is about ten orders higher than for radio frequency (RF) communications, for the same aperture size and transmitted power. Thus, laser communication is the most promising approach to support Internet-like speeds within the constrained size, weight, and power (SWaP) for deep space explorations to Mars, Jupiter, and beyond the solar system[3,4]. However, the communication system has to push many technologies, including transmitted power, telescope aperture, and receiver's sensitivity, to their performance limits to achieve this goal. For instance, the Lunar Laser Communication Demonstration (LLCD) could be considered a remarkable step towards to deep-space laser communication, which achieved 622 Mbps downlink speed, an order of magnitude faster than the best Ka-band radio system flown to the moon (100 Mb/s) on the Lunar Reconnaissance Orbiter in 2009[5]. After LLCD, NASA launched payload of Laser Communications Relay Demonstration (LCRD)[6] in 2021, and is working on a Deep Space Optical Communications (DSOC) experiment which will be carried out on NASA's Psyche spacecraft[7].

Although the recent development of coherent receivers has shown improved sensitivity[8], photon counting-based receiver architecture has competitive efficiency. Additionally, it does not need to be equipped with complex adaptive optics that is indispensable for coherent receivers in the presence of atmospheric turbulence[3]. Furthermore, pulse position modulation (PPM) of photons detected by fast and efficient single-photon detectors (SPDs) is the best choice for deep-space communication where attenuation is huge[9]. Among several existing SPDs, the superconducting nanowire single-photon detector (SNSPD) is preferred to PPM formatted communication due to its high efficiency, low timing jitter and free-running operation mode. The LLCD grounded receiver used four telescopes, each coupled to a four-pixel SNSPD array, to overcome the speed bottleneck[10]. The SNSPD's performance has improved significantly in the years after the LLCD[11]. For instance, the system detection efficiency (SDE) increased to higher than 90%[12–16], and the timing jitter reduced to less than 10 ps[17]. Therefore, a receiver equipped with the state-of-art SNSPDs would exhibit superior performance in terms of sensitivity and speed than its precursor.

In this study, a superconducting nanowire photon-number resolving (PNR) four-quadrant detector (4-QD)-based high-speed laser-communication receiver prototype was built for overcoming some special difficulties in deep-space laser communication. The 4-QD is free space coupled with a maximum system efficiency of 72.7%. After excluding the 0.02 dB, 0.99 dB and 0.25 dB losses from optical window, cryogenic filters and free-space light coupling, respectively, the device detection efficiency was 97.5%. Furthermore, at this bias point, the total dark counts were less than 100 counts per second (cps) and the minimum timing jitter was 78 ps for single-photon detections and 21 ps for six-photon detections. Apart from these advanced performances of SPDs, the receiver has several special features by designing the nanowire electronics elaborately. Instead of using a single nanowire, each pixel of the 4-QD adopts a serial assembly[18,19], where each nanowire is shunted by a resistor, and six shunted nanowires are connected in series. Consequently, this design increases the total photon counting rate to 1.6 Gcps without any additional readout complexity and gives a PNR capability from one to 24 photons simultaneously. At 8-PPM modulation format and using serially concatenated pulse-position modulation (SCPPM) code, the receiver sensitivities were 1.47, 1.76, 3.4, and 7.35 photons/bit at bit rates of 480 Mbps, 800 Mbps, 1.2 Gbps, and 1.5 Gbps, respectively. Additionally, the sensitivity improved by about twofold at a similar bit rate, and the maximum bit rate was more than doubled compared to the LLCD receiver (sensitivity was 3.18 photon/bit at a maximum bit rate of 622 Mbps). Moreover, this serial architecture offered a PNR capability by quantizing the electrical pulse height. The PNR ability overcame the 1-bit dynamic range bottleneck in conventional SNSPDs. Therefore, the receiver in this study was strongly immune to the background photon noise. Experimentally, the receiver maintained an error-free link for the background noise as high as 0.8 photons per PPM slot, indicating its capability of daytime operation in the future. Additionally, the 4-QD offers position sensitivity in a photon counting mode, which is essential for the tracking system to stabilize the downlink beam and is the prerequisite for achieving the best sensitivity. As shown in Fig. 1, building a deep-space receiver with these improvements can be envisioned to provide high-speed bit rates with sensitivity close to the channel capacity limit and assist in tracking the dim beacon light that transmits from a huge distance.

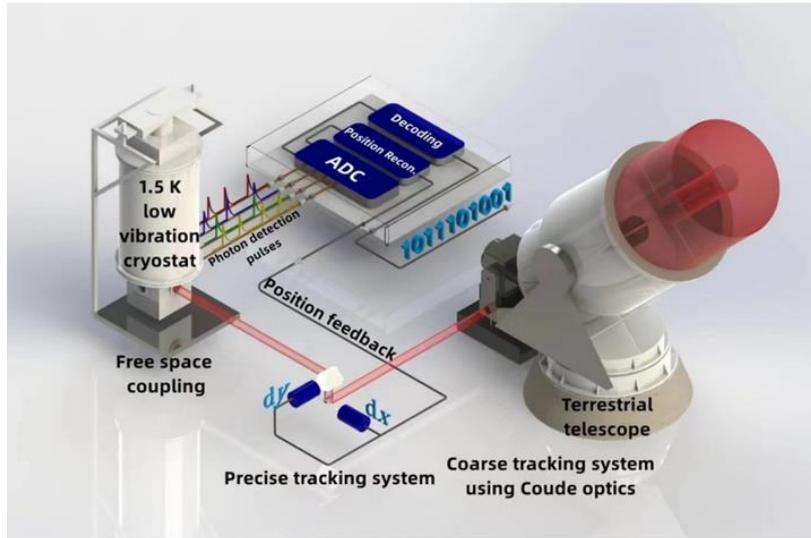

**Fig. 1. Conceptual diagram of the receiver for deep space communications.** The receiver accepts signal light from a Coudé telescope through free-space coupling. Photon detection pulses from the 4-QD mounted in a low-vibrated cryostat are acquired and processed by following electronics at room temperature. Decoded digital bits and beam spot positions are given as output for the users and tracking system, respectively.

## Results

## Detector architecture and performance

The detector must have small timing jitter and fast reset time to reach the Gbps bit rate in PPM format. The state-of-art timing jitter of an SNSPD can be below 10 ps, supporting a theoretical PPM slot frequency of up to 100 GHz. However, the reset time of the SNSPD is the dominant bottleneck due to two tradeoffs. The first tradeoff is between the detector's active area and the reset time. Increasing the active area can improve the optical coupling and the system detection efficiencies. However, a long nanowire is required to cover this active area, giving a large kinetic inductance that slows down the bias current recharging[20]. Although using an array of SNSPDs, which operates multiple pixels in parallel, can shorten the average reset time, a second tradeoff between the readout complexity and the array size is introduced. Individual reading of a large number of pixels increases the thermal load at cryogenic temperature and makes the communication receiver hardware more complicated, including digitization, synchronization, and decoding.

As shown in Fig. 2(a), a 2 × 2 SNSPD array was designed to balance the above tradeoffs. Although the array size was moderate, it could give a position-sensing capability by operating as a photon-counting 4-QD. The array is referred to as a 4-QD in the remaining part of this study to specifically compare its simultaneous communication and tracking abilities with conventional 4-QDs made from semiconductor photodiodes[21]. The nanowire was divided into six segments for each quadrant, referred to as sub-pixels. Each sub-pixel was shunted by a neighboring resistor, designed in a fishbone geometry, as shown in Fig. 2(b). The six sub-pixels were connected in series with a common output port. The equivalent circuit of one such series detector is shown in Fig. 2(c). The serial architecture, which was invented originally for obtaining PNR capability[18], was found to offer a high photon counting rate[19]. As shown in the simulation results in Figs. 2(d) and (e), when one sub-pixel is fired, the recovery time constant is $\tau_f^f = L_k/R_p$, where $L_k$ is the kinetic inductance for each sub-pixel and $R_p$ is the shunted resistance. Although the sub-pixels were connected in series, the fired sub-pixel caused a little reduction in the current through the unfired sub-pixels. Therefore, the six sub-pixels can detect photons successively, equivalently giving a short reset time. However, the whole detector's reset time is relatively slower (recovery time constant is about $6 \times L_k/R_L$), since the amplifier's input impedance $R_L$ is typically 50 Ω, which is far lower than an ideal high impedance[18]. As shown in Fig. 2(e), the output pulses have a strong pile-up effect at a high counting rate. However, digital signal processing can overcome this pile-up effect to recover the arrival time and height of the pulse, which is discussed later in the communication section.

The nanowire was made from a 5 nm thick niobium nitride (NbN) film. NbN has lower kinetic inductivity and faster thermal cooling than amorphous films, such as WSi and MoSi[22]. The array covered an area of 20 μm × 20 μm. The nanowire was 90 nm wide and 60 μm long for each sub-pixel, giving $L_k$ = 52 nH. The shunt resistance was $R_p$ = 50 Ω, pushing the reset time constant $\tau_f^f$ to 1 ns for each sub-pixel. This time constant was just above the latching limit[23] to maximize the detection speed. Shunt resistors were fabricated as close as possible to the sides of the nanowires to avoid any parasitic inductance from connection wires. However, it was found that at a high counting rate, the Joule heating of a straight wire resistor increased the local temperature, making the detector latch at a high bias current. Thus, the resistors were designed into a fishbone geometry to speed up the cooling. As shown in the simulation result in Fig. 2(b), the maximum local temperature was only raised to 2.5 K at an operation temperature of 1.5 K with the fishbone geometry (simulation details are described in the supplementary materials).

The total counting rate of the array was characterized, and the results are shown in Fig. 2(f). As each pixel can output pulse height proportional to the incident photon number and the coherent laser pulse has a Poisson distribution for photon numbers, there are two metrics to characterize the counting speed, which are the photon counting rate $CR_{\text{phn}}$ and pulse counting rate $CR_{\text{pls}}$. The laser repetition rate was set as 1 GHz, and the attenuation was swept to vary the incident photon flux intensity. Furthermore, at strong attenuations, the mean photon number per pulse $\mu$ was much less than one. Therefore, the incident photons were almost in pure single-photon events, and $CR_{\text{phn}}$ was almost equal to $CR_{\text{pls}}$. As $\mu$ increased to have more multi-photon events, $CR_{\text{phn}}$ was higher than $CR_{\text{pls}}$. When the photon and pulse detection efficiencies dropped by 3 dB, the corresponding counting rates were $CR_{\text{phn}}^{\text{3dB}} = 1.61$ Gcps and $CR_{\text{pls}}^{\text{3dB}} = 1.40$ Gcps. In a communication system, $CR_{\text{pls}}$ determines the maximum repetition rate of the PPM symbols, and $CR_{\text{phn}}$ benefits the forward error correction (FEC) performance for a more precise calculation of the likelihood ratio (LLR). Such performance will be discussed later.

The timing jitter of the first quadrant pixel is shown in Fig. 2(g), since the detector has PNR capability. The timing jitter of the single-photon and six-photon response output pulses were 78 ps and 21 ps, respectively. The difference was primarily caused by the voltage noise, adding additional jitter for single-photon responses with a relatively low signal-to-noise ratio (SNR). The jitter data of the other three quadrants are given in the supplementary material. Although these values were below the state-of-art values, they supported PPM slot frequencies up to 10 GHz and bit rates over 1 Gbps. Furthermore, it must be addressed that all the detector performances are in tradeoffs, and this study seeks an optimal solution for communications.

Furthermore, the detector was integrated with an optical cavity to obtain high light absorption. A three-axis cryogenic piezo nano-positioner was installed, and a high NA aspheric lens was mounted on it to have the best coupling efficiency. The focused Gaussian beam was 19.2 μm wide, defined at the 1/e² points. As shown in Fig. 2(i), the maximum system detection efficiency was 72.7% by scanning the spot position. Excluding the optical window, cryogenic filtering, and free space light coupling losses of 0.02dB, 0.99 dB, and 0.25 dB respectively, the device efficiency was 97.5% (loss calibrations are given in the supplementary materials). The characterization results of the SDEs of the four quadrants are shown in Fig. 2(h). During individual characterization, the coupling efficiency was reduced to 56.1% since the active area of one quadrant was smaller than the spot size. Therefore, the average system detection efficiency of one quadrant was 42.5%. The noticeable saturation plateau in the photon counting vs. bias current curve indicated that the internal quantum efficiency of the nanowire was near unity.

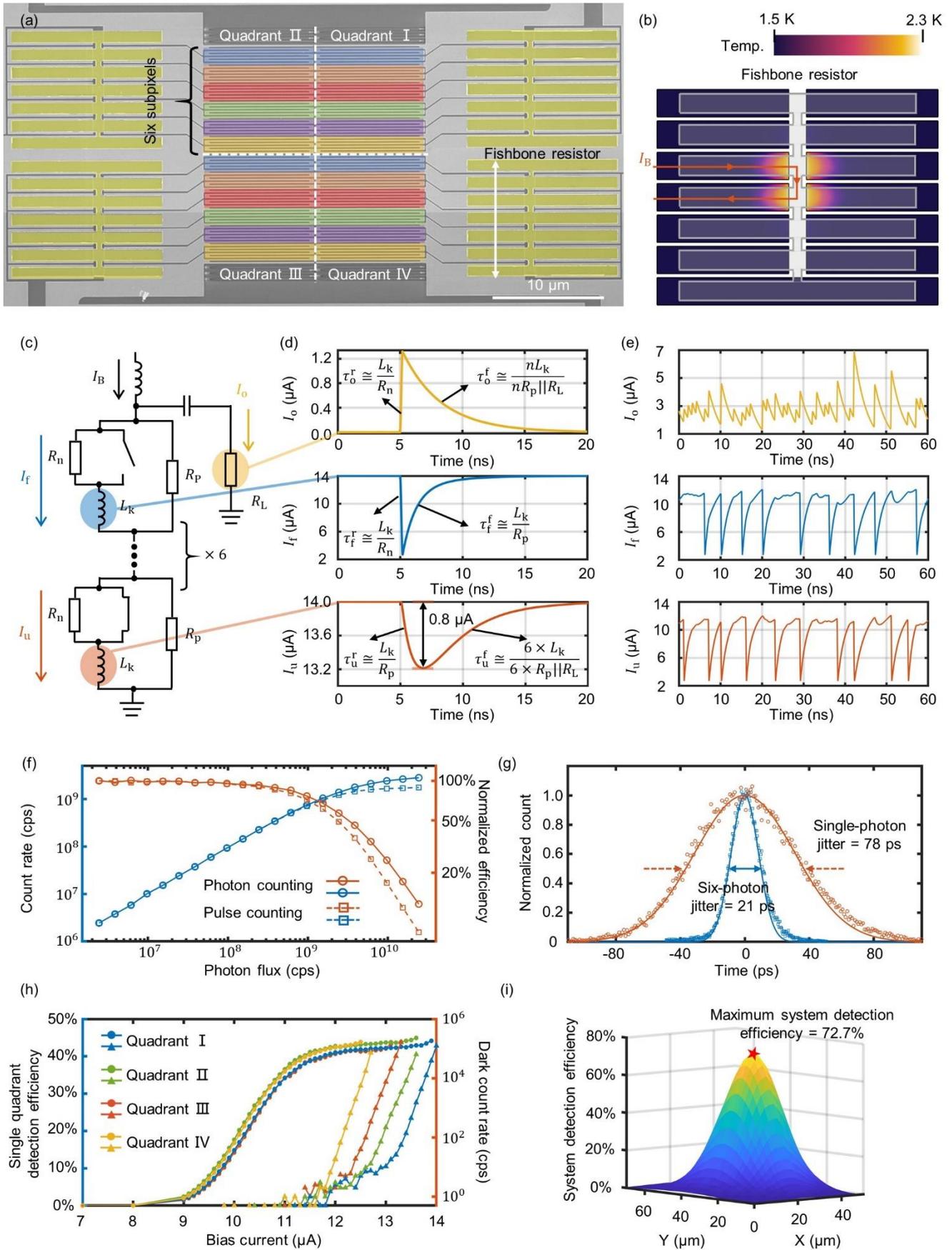

**Fig. 2. Detector architecture and performance. a.** SEM image of the 4-QD made from superconducting

nanowires. **b.** Simulation of the temperature distribution of the NbN surface below the "fishbone" resistor when the third sub-pixel fires. **c.** The equivalent circuit of one pixel was designed in the serial architecture. **d.** and **e.** The transient currents of the output $I_o$, fired sub-pixel $I_f$, and unfired sub-pixel $I_f$ at single detection and high counting speed, respectively. **f.** Dependence of the total photon and pulse counting rates on the incident photon number. The corresponding normalized detection efficiency is plotted on the right y-axis. **g.** Timing jitters of single-photon and six-photon responses for the first quadrant. **h.** System detection efficiency vs. bias current for all four pixels. **i.** The total system detection efficiency vs. the center position of the focused light spot.

## Receiver sensitivities at different bit rates

A PPM format communication testbed was built with the superconducting nanowire 4-QD, and the sensitivity of the receiver prototype at different bit rates was characterized, as shown in Fig. 3(a). The system consisted of a transmitter, a free-space channel with tunable attenuation, and a receiver to mimic the deep-space communication scenario. The communication used the 1/2 rate SCPPM[24], which had been demonstrated successfully in the LLCD. A detailed description of the set-up is given in the method section. The detection pulses piled up at high modulation rates, similar to the inter-symbol interference problem in conventional communication systems. Although this was the evidence of fast photon counting, it made difficult for the decoder to get the correct photon information. A signal pre-processing strategy, including a matched filter, deconvolution, and peak searching, was introduced to overcome this problem. An example is shown in Fig. 3(b), and its details are given in the method section. After the pre-processing, photon arrival times and detection photon numbers were extracted, which input to the SCPPM encoder for correcting errors.

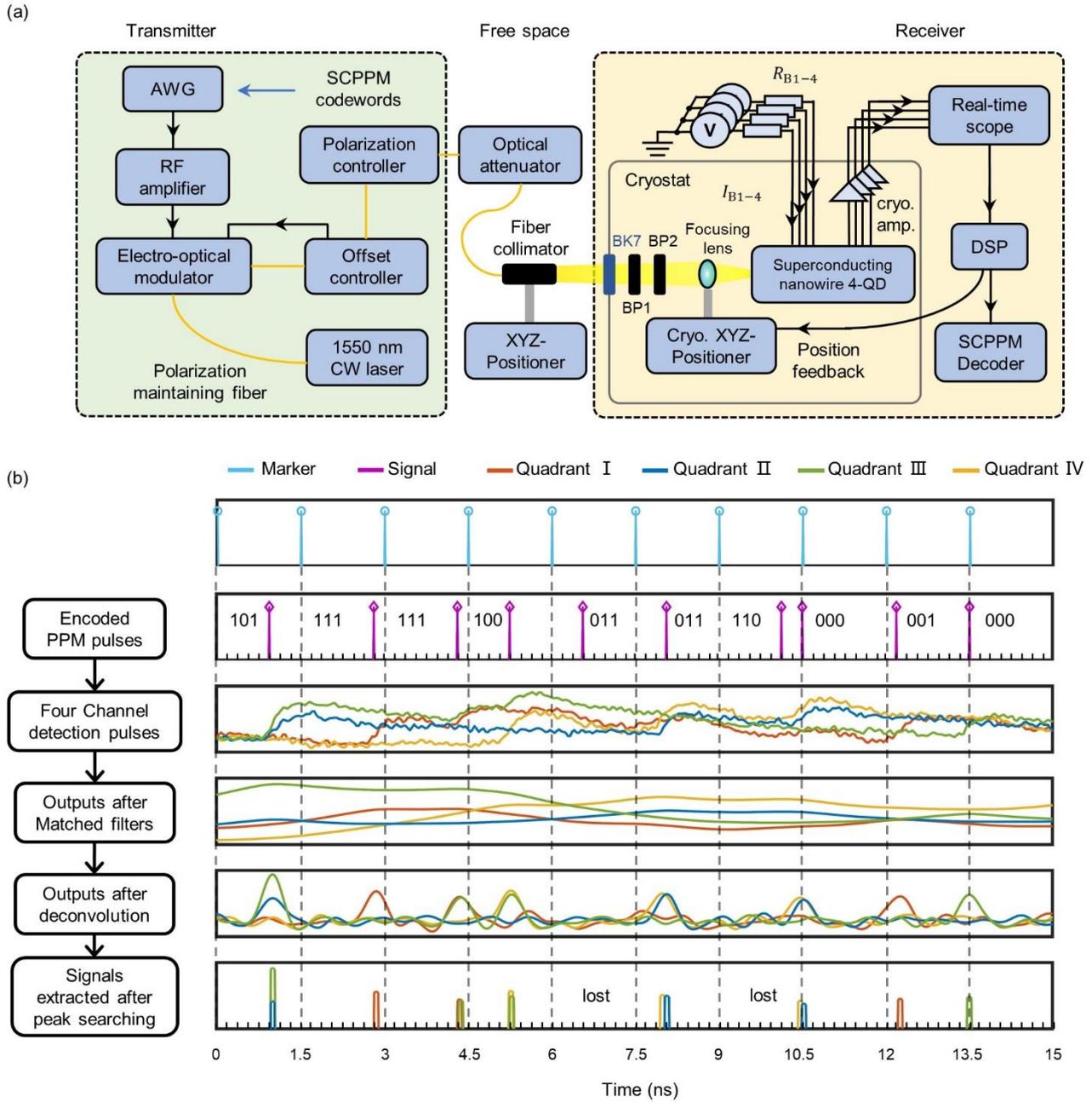

**Fig. 3. Communication set-up and signal pre-processing. a.** Set-up diagram for the communication testbed. **b.** Illustrations of signal pre-processing steps. M-PPM has an order of M = 8 with slot width of 187.5 ps, resulting in a bit rate of 1 Gbps, including the 1/2 encoding rate.

The 8-PPM format was chosen to maximize the bit rate due to the short reset time of the 4-QD. The bit rates were varied from 120 Mbps to 1.5 Gbps (channel bit rates before FEC were from 240 Mbps to 3 Gbps) by adjusting the slot width. No dead time slot was inserted between the adjacent PPM symbols. The number of sending bits was $10^5$. The light spot was enlarged to a $1/e^2$ width of 28.4 μm to take full advantage of all the sub-pixels, reducing the total system efficiency to 57.5%. The measured bit error rates (BERs) at different bit rates $R$ are shown in Figs. 4(a) and (b). The incident photon number per pulse (*IPN*) and the detected photon number per pulse (*DPN*) were calibrated for the x-axes in Figs. 4(a) and (b), respectively. The SCPPM error correction was effective compared to the hard decision. The BER dropped rapidly to zero as the photon intensity increased. By extracting the photon levels where $BER < 10^{-5}$ as the photon number thresholds, dependences of *IPN* thresholds and *DPN* thresholds at different $R$ can be obtained, as shown in Fig. 4(c). Meanwhile, the system detection efficiency defined by the ratio between *IPN* and *DPN* was plotted. Similar to the photon counting rates shown in Fig. 2(f), the system detection efficiency on average reduced as $R$ increased. Therefore, the receiver required a higher *IPN* to support an error-free communication. *DPN* thresholds showed less influence on $R$, indicating that the timing jitter and voltage noise deteriorated the sensitivity at high bit rates.

The sensitivity (in unit photons per bit) characterized the minimum required photons per receiving one bit and the spectral efficiency (in unit bits per second per Hz) characterized the communication rate at a given modulation bandwidth, i.e., the reciprocal of the PPM slot width. These values were calculated to compare with previous demonstrations. As shown in Fig. 4(e), the PPM format, in general, shows excellent sensitivity but poor spectral efficiency. However, as shown in Fig. 4(d), this study demonstrated better sensitivity and a higher maximum bit rate than the reported results of the LLCD receiver. The LLCD receiver had a sensitivity of 3.18 photon/bit at 622 Mbps and spectral efficiency of 0.125 bits/sec/Hz. The results in this study were 1.47 photons/bit at 480 Mbps and 1.76 photons/bit at 800 Mbps. Furthermore, by sacrificing some detection efficiency and increasing the PPM slot rate to 8 GHz, a 1.5 Gbps bit rate could be obtained with a sensitivity of 7.41 photons/bit. Additionally, this high bit rate performance is competitive to the DPSK coherent receiver (7.3 photons/bit at 1.244 Gbps) in the LCRD experiments[25,26]. Therefore, these results are encouraging and show that a photon-counting basic communication system can continue to improve for better sensitivity and faster bit rates by improving the performance of detectors.

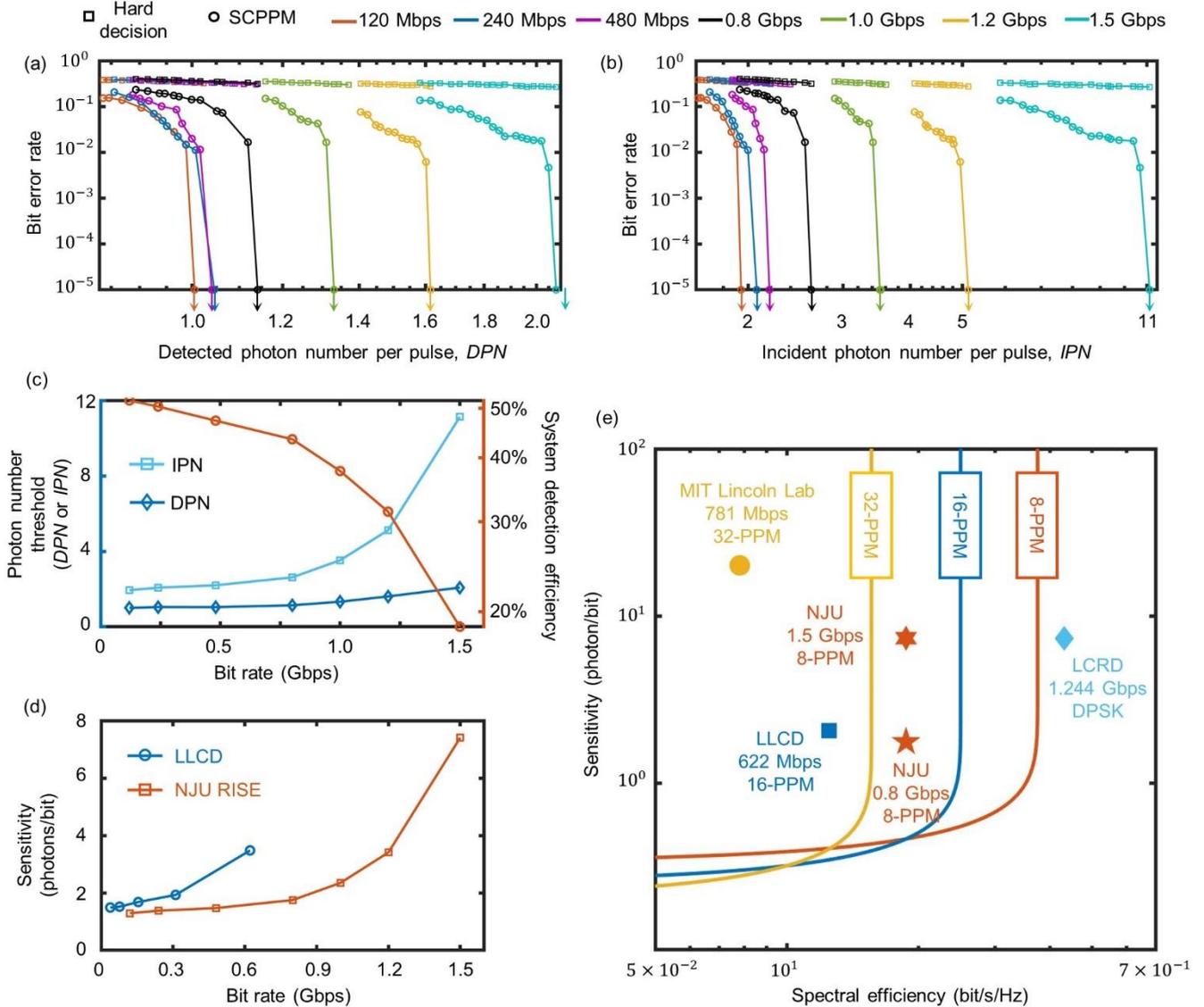

**Fig. 4. Communication results. a.** and **b.** Bit error rate (BER) vs. detected and incident photon numbers at different bit rates, respectively. Arrows indicate that at these photon number thresholds, BER drops to zero for all the decoded bits ($10^5$). **c.** Dependences of the photon number thresholds of *DPN*, *IPN*, and the system detection efficiency at different bit rates. **d.** Dependences of the sensitivity on bit rate for LLCD and results of this study (NJU RISE). **e.** Dependences of the sensitivity on spectral efficiency. Data from several related projects are plotted with the results of this study for comparison. The lines are theoretical curves for PPM format communications.

# Communications at strong background photon noise

Typically, laser communications are preferred to be operated at night due to a clear background. However, daylight operations are welcomed because the total communication duration can be increased, and for certain missions where the transmitter can only be seen on Earth during daylight. Particularly, for solar system boundary exploration, observation and communication have to be done during the daytime to face the sun. However, the strong background noise from the daylight sky, which is prominent (for instance, 0.78 photons per PPM slot as calculated in Ref [3]) compared with the signal photons received from very far away, prevents the operation of laser communication during the daytime.

As mentioned above, the series nanowire architecture has a quasi-PNR capability. As shown in Fig. 5(a), the output pulse height is proportional to the number of fired nanowires, from which the photon-number distribution with a known spatial distribution of the incident light spot can be deduced. As shown in Fig. 5(b), the sum of all four outputs gives a maximum photon number of 24. Furthermore, this is important since such PNR capability provides a dynamic range of 4.6 bits instead of 1 bit, bridging the gap between an SPD and a linear photodetector. Therefore, for a soft-decision FEC, such as SCPPM, the detected photon number will convert into an accurate *LLR* based on an estimate of the channel[27] as follows:

$$LLR = \log_2 \frac{p(k|1)}{p(k|0)} = k \cdot \log_2\left(1 + \frac{n_s}{n_b}\right) - n_s \qquad (Eq.1)$$

where $k$, $n_s$, and $n_b$ represent the detected photon number, the average signal photon number, and the average background noise photon number, respectively. As shown in Fig. 5(c), compared to single-photon counting mode where $k$ was in a binary mode ($k = 0$ for no detection and $k = 1$ photon detections independent of photon number), a PNR capability gave a gain of 3.3 dB in the incident photon number threshold at a background noise of 0.2 photons/slot and bit rate of 800 Mbps.

The background photon noise was swept, and the incident signal photon thresholds at two different bit rates of 120 Mbps and 800 Mbps were extracted. The results are shown in Figs. 5(d) and (e). As the background noise increases, more errors caused by noise photon detections appear. Therefore, a higher incident photon number was required for the FEC. There was a maximum background noise level that the FEC could not tolerate and stopped working. The maximum noise levels for single-photon detections were about 0.1 photons/slot at 120 Mbps and 0.2 photons/slot at 800 Mbps. However, the maximum noise levels for PNR detections increased to about 0.8 photons/slot at 120 Mbps and 0.4 photons/slot at 800 Mbps.

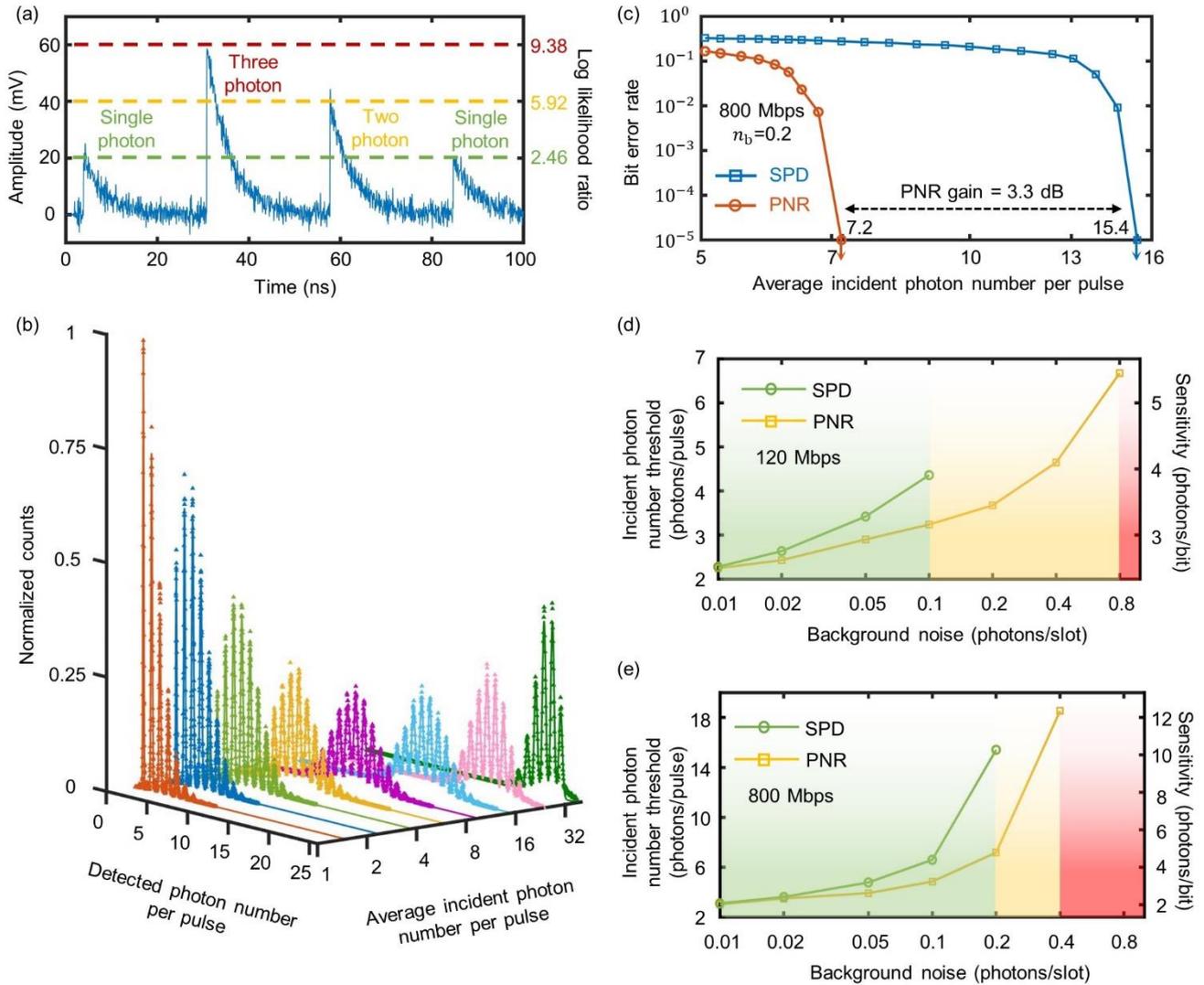

**Fig. 5. Communications at strong background photon noise. a.** Examples of output pulse waveforms for different photon number detections and corresponding *LLR* values. The estimated background photon noise is $n_b = 0.1$. **b.** Photon number distributions detected by the 4-QD at different incident photon intensities. A matched filter is applied to improve the discrimination between adjacent peaks. **c.** Error correction performance comparison of FEC, with and without PNR detections. The background noise is $n_b = 0.2$ photons/slot. **d.** and **e.** Dependences of incident photon number thresholds for FEC, with and without PNR detections at bit rates of 120 Mbps and 800 Mbps, respectively. The green and yellow areas mark the range where the FEC can work.

The red areas mark the range where the background noise is so strong that the FEC fails.

## Beam position sensing by the photon-counting 4-QD

Accurate beam pointing is a prerequisite for stabilizing a laser communication link, which needs an accurate acquisition, tracking, and pointing (ATP) system. For deep space communications, huge distance causes a serious attenuation for the beacon beam and an ATP system may use only uplink beacon beam or a beaconless solution due to the long one-way light travel latency[28]. However, fine tracking of the signal beam is always necessary on the ground receiver to maximize the sensitivity and cancel spot drifting caused by vibrations of the platform or atmosphere turbulence. Therefore, a deep space receiver equipped with a single-photon position sensing detector for beam tracking is a promising technique[29].

A 4-QD, which has a 2 × 2 array configuration, is a widely used simple position sensor for acquisition and tracking. As shown in Fig. 6(a), when the beam spot was scanned, the photon counting rates $CR_1$, $CR_2$, $CR_3$, and $CR_4$ from the four quadrants of the superconducting nanowire 4-QD had clear position dependences. The maximum likelihood estimator of the incident flux intensity $\Phi$ is given by $\widehat{\Phi} = h\nu \cdot CR$, where $h\nu$ is the photon energy. Therefore, a basic 4-QD centroid algorithm by replacing the inputs to $CR_1 - CR_4$[30] can be applied, which are as follows:

$$\Delta x = k_x \frac{(CR_1 + CR_4) - (CR_2 + CR_3)}{CR_1 + CR_2 + CR_3 + CR_4}$$
$$\Delta y = k_y \frac{(CR_1 + CR_2) - (CR_3 + CR_4)}{CR_1 + CR_2 + CR_3 + CR_4}$$

where $k_x$ and $k_y$ are the coefficients for converting the counting rate differences into position offsets $\Delta x$ and $\Delta y$ from the centroid of the beam spot, respectively. As shown in Figs. 6(b) and (c), the calculated $\Delta x$ and $\Delta y$ show nearly linear dependences to spot centroids within the range of the total area of the nanowire. The root mean square errors (RMSEs) for $\Delta x$ and $\Delta y$ are 0.011 and 0.021, respectively. Therefore, these results demonstrate the expected position-sensing ability of the superconducting nanowire 4-QD.

The position-sensing ability can be integrated simultaneously with the communication to prevent the splitting of the incident signal light into a second optical path, improving the sensitivity of the receiver at the system level. As shown in Fig. 6(d), when the beam spot shifts outside the center of the detector, the counting rates vary significantly. Therefore, accurate focusing of the beam spot at the center can divide the incident photon flux evenly on every detector, maximizing the detection performance of all pixels.

The discussions above suggest that the communication *BER* is related to the beam position. As shown in Figs. 6(e) and (f), the *BER* was measured by scanning the beam spot positions at two different incident photon intensities. The *BER* rolled down rapidly when the beam spot approached the center of the detector. Consequently, a spot shifting margin $\varphi_{BER<10^{-5}}$ below which the *BER* was less than $1 \times 10^{-5}$ was defined. $\varphi_{BER<10^{-5}}$ was 12 μm at bit rate of 800 Mbps and *IPN* = 5.3 photons/slot. Considering the detector had a total detection area of 20 μm × 20 μm, it was indicated that until the laser beam was not shifted off the center by 60%, an error-free communication could be maintained. However, when the incident photon number reduced to 3.2 photon/pulse, $\varphi_{BER<10^{-5}}$ reduced to 4 μm. Therefore, these results demonstrate that an accurate centering of the beam spot is necessary for obtaining the best sensitivity.

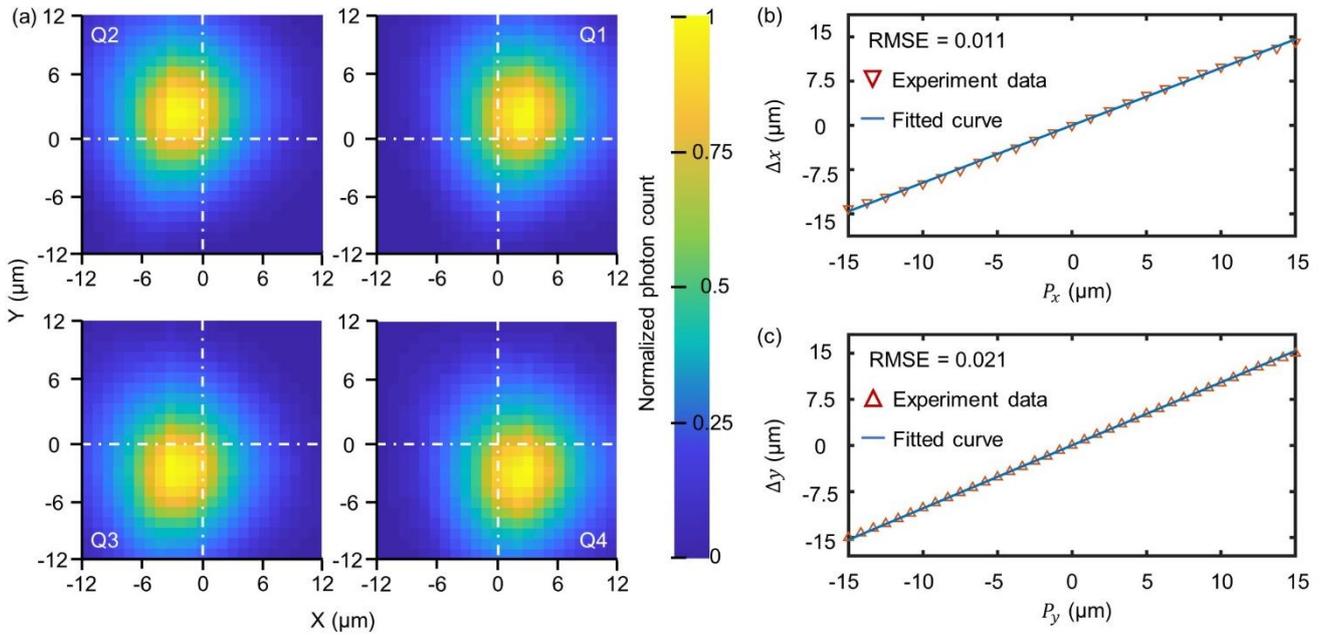

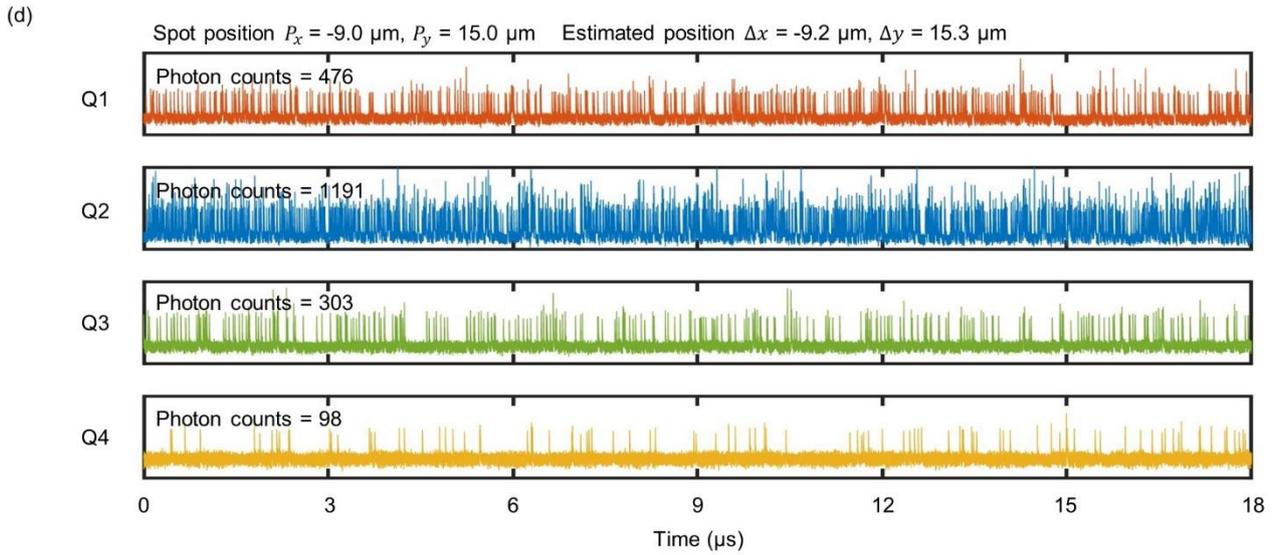

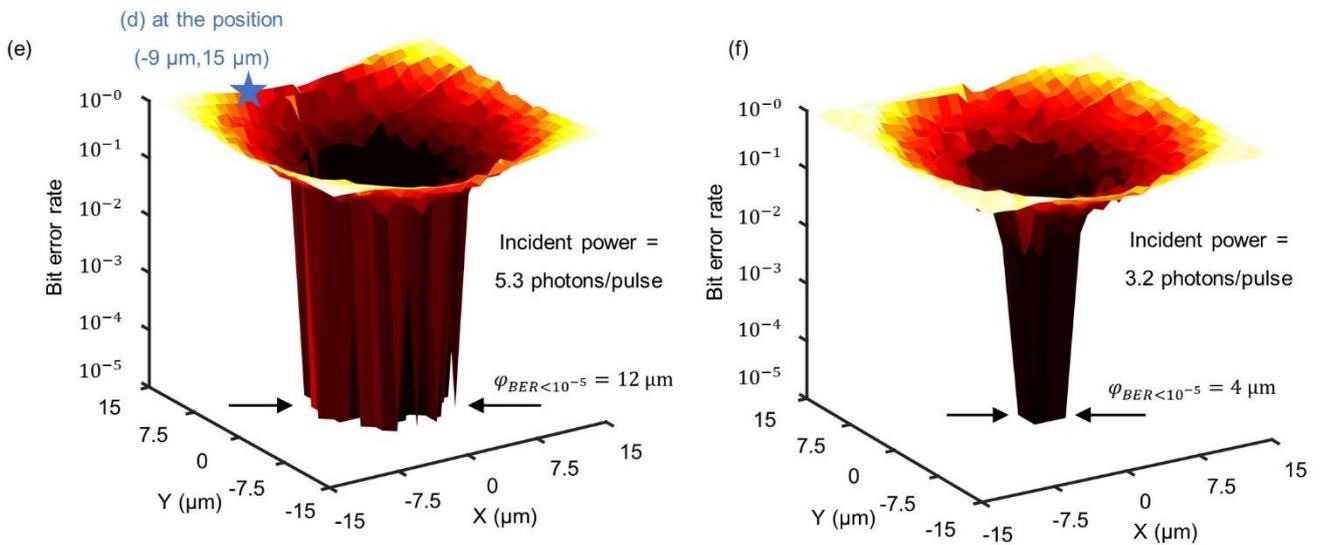

**Fig. 6. Beam position sensing by the photon-counting 4-QD. a.** Photon counting rate maps for the 4-QD by scanning the beam centroid. **b.** and **c.** Estimated position shifts $\Delta x$ and $\Delta y$ based on the counting rate differences from the 4-QD, respectively. The lines are fitting curves by using error functions. **d.** Output waveforms from each pixel of the 4-QD during 8-PPM format communication at a bit rate of 800 Mbps and position of (-9.0 μm, 15.0 μm). **e.** and **f.** Communication *BER* vs. position at incident photon numbers 5.3 photons/pulse and 3.2 photon/pulse, respectively. The star marks the position where waveforms shown in (d) are taken.

# Conclusion

The experimental results showed that the proposed receiver prototype addressed several challenges in deep-space laser communications through joint development of the detector, cryogenic engineering, and signal processing methods. With a small number of output lines, bit rates from 120 Mbps to 1.5 Gbps were demonstrated, covering a broad range for various space laser communication missions. Benefiting from the PNR capability, the receiver was robust to background photon noise and exhibited daylight operation potential. Meanwhile, the 4-QD architecture gave a beam tracking capability simultaneous with communications at photon-counting mode, which offers direct tracking of the signal beam and stabilizes the receiver at its best sensitivity. With these specialties mentioned, the authors envision that the receiver prototype would find applications in future deep-space missions. The future tasks include hardware implementation of signal processing, decoding and tracking, light coupling from a large telescope to a cryogenic device, and further optimizing the detector's performance.

# Methods

## Communication set-up description

A high-speed arbitrary wave generator with a sampling rate of 64 GHz was used for generating the coded PPM electrical pulses at different bit rates. Then, the electrical pulses were amplified and input to an electro-optical modulator (EOM) for generating optical pulses. A continuous laser at a wavelength of 1550 nm was used as the seed source for the EOM. A bias offset controller was used to stabilize the DC offset of the EOM to maximize the extinction ratio of the pulsed laser, which could be 40 dB. A polarization controller was used to tune the light polarization to maximize the detection efficiency. Finally, an optical attenuator was used to adjust the power to weak signals for simulating the loss in deep space.

Furthermore, the attenuated optical signal was collimated into free space at room temperature. A high NA (NA = 0.54) aspheric lens was placed to focus the light onto the detector. A coarse XYZ and a fine cryogenic piezo positioner were used to align and focus the light spot. The light passes through a K9 glass window, a 1750 nm short-pass filter, and a 1550 nm band-pass filter with a total loss of 1.01 dB. The two filters were placed at 4 K to remove the background photon noise from thermal radiation. Additionally, four cryogenic amplifiers were placed at 4 K for individually amplifying the four pixels in the 4-QD. The detection pulses were acquired by a real-time oscilloscope with a maximum sample rate of 20 Gsps and then processed offline, including a series of signal pre-processing steps and the SCPPM FEC.

## Signal pre-processing steps

The 4-QD had four outputs $n = 1 – 4$. Each output pulse $r_n(t)$, the superposition of the signal $s_n(t)$ and noise $g_n(t)$, is as follows:

$$r_n(t) = s_n(t) + g_n(t) \tag{2}$$

The photon detection pulse $s_f(t)$ can be expressed by a double exponential function as follows:

$$s_f(t) = A \cdot (e^{-t/\tau_1} - e^{-t/\tau_2}) \tag{3}$$

where, $\tau_1$ and $\tau_2$ represent the rising and falling times of the pulse, respectively. $A$ represents the pulse amplitude of single photon response. First, a matched filter to maximize the signal-to-noise ratio described in our previous study was used[31]. $y_n(t)$, the waveform after matched filter, can be written as follows:

$$y_n(t) = r_n(t) \otimes s_f(t_0 - t) \tag{4}$$

where $\otimes$ denotes the convolution operation. However, the pulses were still indistinguishable after the matched filter, so a deconvolution method was needed to identify their positions and height. The deconvolution model $D(t)$ is as follows:

$$D(t) = s_f(t) \otimes s_f(t_0 - t) \tag{5}$$

Then, the deconvoluted signal $V$(t) is as follows:

$$V(t) = y_n(t) \odot D_t(t_0 - t) \qquad (5)$$

where $\odot$ denotes the deconvolution operation. The final step was to find the peak locations and heights of $V(t)$, marking the photon arrival times and detected photon number, respectively.

(2021).

## Data availability

The data that support the plots within this paper and other findings of this study are available from the corresponding author upon reasonable request.

## Acknowledgements

We thank the other RISE members for assistance in nanofabrication, measurements and providing instruments. We appreciate colleagues and friends who gave us professional comments and discussions during this project. This work was supported by the National Natural Science Foundation (nos. 62227820, 62071214, 61571217 and 11227904), the Program for Innovative Talents and Entrepreneur in Jiangsu, the Innovation Program for Quantum Science and Technology (No. 2021ZD0303401), the Fundamental Research Funds for the Central Universities, the Priority Academic Program Development of Jiangsu Higher Education Institutions (PAPD) and the Jiangsu Provincial Key Laboratory of Advanced Manipulating Technique of Electromagnetic Waves.

## Author contributions

H.H. and Q.-Y.Z. conceived the idea. H.H. fabricated the device and implemented the experiments. H.H. and Q.-Y.Z. performed the numerical simulations. Y.-H. H., H.W., J.-W. G., S.C., X.-C.T., L.-B.Z., X.- Q.J., and L.K., Wang helped with device fabrication. J.D., Z.L., S.-Y. R., Y.-J. Z., S.-H. W., C.W., H.L., and J.C. helped with the cryogenic setup. Z.-J. L. helped with picture drawing. H.H. and Q.-Y.Z. discussed the results and wrote the manuscript. Q.-Y.Z. and P.-H.W. supervised the work.

## Corresponding author

Correspondence to Qing-Yuan Zhao   qyzhao@nju.edu.cn

## Competing interests:

H.H., Q.-Y.Z. and L.K. applied for a Chinese patent (no. 2019105004548). The remaining authors declare no competing interests.

24 / 24